\documentclass[reprint,amsmath,amssymb,aps,pra]{revtex4-2}

\usepackage{graphicx}  
\usepackage{color}
\usepackage{dcolumn}
\usepackage{bm}

\begin{document}

\preprint{AIP/123-QED}

\title{Interaction Quench Dynamics and Stability of Quantum Vortices in Rotating Bose-Einstein Condensates}
\author{L. A. Machado$^{1}$, B. Chatterjee$^{2}$, M. A. Caracanhas$^{1}$, L. Madeira$^{1,3,4}$, V. S. Bagnato$^{1,5}$, B. Chakrabarti$^{1,6}$ }
\email{barnali.physics@presiuniv.ac.in}
\affiliation{$^1$Instituto de Física de São Carlos, Universidade de São Paulo, CP 369, 13560-970 São Carlos, SP, Brazil.}
\affiliation{$^2$Department of Physics, Asansol Girls’ College, Asansol, 713304, West Bengal, India}
\affiliation{$^3$ INFN-TIFPA Trento Institute of Fundamental Physics and Applications, Via Sommarive 14, I-38123 Trento, Italy}
\affiliation{$^4$ European Centre for Theoretical Studies in Nuclear Physics and Related Areas (FBK-ECT*), Strada delle Tabarelle 286, Trento, Italy}
\affiliation{$^5$ Department of Biomedical Engineering, Texas A\&M University, College Station, Texas 77843, USA}
\affiliation{$^6$ Department of Physics, Presidency University, 86/1   College Street, Kolkata 700073, India.}
\date{\today} 

\begin{abstract}
We theoretically investigate the non-equilibrium dynamics of quantum vortices in a two-dimensional rotating Bose-Einstein condensate following an interaction quench. Using an {\it ab initio} and numerically exact quantum many-body approach, we systematically tune the interplay between interaction strength and angular velocity to prepare quantum vortices in various configurations and examine their post-quench dynamics. Our study reveals distinct dynamical regimes: First, vortex distortion accompanied by density cloud fragmentation, matching the initial vortex number and second, vortex revival, where fragmented densities interact and merge. Notably, we observe complete vortex revival dynamics in the single-vortex case, pseudo-revival in double and triple vortex configurations, and chaotic many-body dynamics in systems with multiple vortices. Our results reveal a universal out-of-equilibrium response of quantum vortices to interaction quenches,  highlighting the importance of many-body effects with a possible exploration in quantum simulation with ultracold quantum fluids.
\end{abstract}


\maketitle

\section{Introduction} \label{sec:intro}

Since the discovery of superfluidity in $^4$He, it is known that mechanical rotations of the superfluid give rise to quantized vortices \cite{donnelly1991}. These topological defects reconcile the irrotational flow, and the quantization of circulation, given in units of $h/m$, where $h$ is the Planck constant and $m$ is the mass of the particle \cite{onsager1949,feynman1955}. Following the realization of Bose-Einstein condensation of trapped atomic clouds~\cite{Dalfovo,Dalfovo1,Dalfovo2}, precise generation of quantum vortices in different trap geometries has been achieved by laser stirring~\cite{Inouye}, rotating magnetic traps~\cite{Madison,Madison1} and oscillatory excitations \cite{seman2010}. Vortices are also observed in experiments using quantum engineering techniques based on atom-field coupling~\cite{Leanhardt} and topological phase manipulation~\cite{Lin}. The study of vortex dynamics has been pivotal in understanding superconductivity, superfluidity, and nonlinear optics~\cite{Zwierlein,Blatter}.

In particular, the investigation of rotating Bose-Einstein condensates (BECs) has emerged as a central topic in the study of quantum vortices due to their interesting features, which include an array of orderly aligned vortices in the lowest Landau level limit \cite{ho2001,baym2005,Schweikhard}, the quantum phase transition to highly correlated ground states in the limit of higher rotation rates of the Bose gas, also referred to as quantum Hall states \cite{feder2007}, Tkachenko oscillations \cite{Coddington} as well as the bending and reconnection of vortex lines \cite{ferrari2017}. A rapidly rotating condensate in a harmonic trap generates a periodic array of vortices that form a triangular lattice \cite{Fetter}. In contrast, rotating BECs confined in optical lattices undergo structural transformations \cite{Williams}. Furthermore, BECs subjected to rotating asymmetric traps \cite{engles2002}, quartic-quadric traps and ring geometries also exhibit unique and intriguing features \cite{Brito,Ramanathan}.

The mean-field Gross-Pitaevskii (GP) equation is often the framework of choice for understanding vortex geometry and dynamics in BECs. The mean-field theory generally applies to systems with a large number of particles and weak interactions, where all particles are assumed to be coherent and occupy a single condensed state. Consequently, the analysis focuses on the single eigenvalue and eigenstate of the reduced one-body density matrix. In the rapid development of quantum gases experiments~\cite{Nir,Gauthier,Kwon,Shaun,Oliver,Chomaz,Tajik,Matthew}, superfluid atomic gases in highly tunable systems offer a platform to investigate vortex dynamics in regimes of strong interaction and ultra-fast rotation. In these regimes, the single-orbital mean-field theory may fail to capture quantum correlations, and nonlinear effects arising from mean-field interactions may not correspond to the correct description of the physical system. Even when mean-field vortices resemble the ones of the {\it ab initio} many-body solution of the Schr\"odinger equation, many-body features are not captured by the mean-field theory~\cite{many-mean}. Apart from failure in capturing quantum correlations for rapidly rotating condensates, the GP equation proves inadequate to describe the mesoscopic system, a finite ensemble of a few tens to hundreds of ultracold atoms~\cite{meso1,meso2}. In such finite-size systems, quantum fluctuations become enhanced due to a stronger finite-size effect. 

In this context, one must employ a correlated many-body anstaz, which can handle quantum correlations arising from the strong interplay between rotation and strong interaction. In this work, we use the multiconfigurational time-dependent Hartree (MCTDH) method to account for fragmentation, the hallmark of MCTDH. We exhibit how the concept of fragmentation helps gauge the validity of the mean-field description and the need for many-body methods. Fragmentation refers to the situation where the total many-body wave function of an interacting system is spread across multiple single-particle orbitals rather than being predominantly concentrated in a single state, as is the case of the mean-field approach. This concept captures the idea that the system deviates from the condensate regime, where most particles occupy a single state, and enters a regime where multiple quantum states are occupied. Significant fragmentation can occur in the presence of strong correlations \cite{lode23}, complex trapping potentials \cite{gerbier21}, or nonequilibrium dynamics \cite{lamporesi18}.

Fragmentation is also present in the description of quantum droplets, self-bound states formed by a delicate balance between attractive and repulsive interactions in a many-body quantum system, where we need to depart from mean-field approaches to properly account for quantum correlations~\cite{Cabrera,Semeghini,Chomaz1}. The collapse originating from a strong mean-field attraction is stabilized by quantum fluctuations, thus directly manifesting beyond-mean-field effects~\cite{Petrov1,Petrov}. The beyond-mean-filed effects are also observed in rotating quantum droplets with embedded vorticity, which exhibit distinct instability behavior~\cite{Dong,Chen}.

The impact of fragmentation on the breaking of the ground state density of rotating BECs in the weakly interacting limit and confined in various anharmonic trap geometries such as an elongated trap, a three-fold symmetric trap, and a four-fold symmetric trap has been thoroughly studied by employing Multiconfigurational Time-Dependent Hartree for Bosons (MCTDHB)~\cite{Dutt,Dutt1,budha_vortex,roy2024rotationquenchestrappedbosonic}. 

Although the ground state properties of quantum vortices are well-understood at the mean-field level, and much progress has been made in many-body approaches applied to strongly interacting systems~\cite{Ortiz1995,Giorgini1996,Vitiello1996,Madeira2016,Madeira2017,Madeira2019}, the out-of-equilibrium dynamics remains challenging. A simple protocol for probing exotic nonequilibrium dynamics is a quench, which consists of preparing the initial setup in the ground state of a given Hamiltonian and then suddenly varying one of its parameters to a different value. The dynamical correlations that arise during the quenching process add more complexity, limiting ongoing research in this area~\cite{Dutt,Dutt1,budha_vortex,roy2024rotationquenchestrappedbosonic,banerjee2024collapsequantumvortexattractive}.

In this work, we advanced the study of two-dimensional (2D) vortex dynamics beyond the conventional mean-field approaches. We consider a gas of atoms with short-range interactions confined inside a 2D disk. We investigate vortex configurations in the strongly interacting regime and across slow-to-fast rotation scenarios, analyzing their stability under interaction quenches. To explore beyond-mean-field effects, we compute the numerically exact solution of the many-body Schr\"odinger equation using the MCTDHB, implemented via the MCTDHX package ~\cite{MCTDHX}. We consider $N=100$ atoms interacting with pairwise contact interactions modeled as narrow Gaussians and trapped in a hard-wall disk. The number of particles may seem untypical in the context of usual BEC experiments; however, fragmentation is shown to decay as $1/\sqrt{N}$~\cite{Meso}. Thus, relatively small particle numbers are conducive to exploring the beyond-mean-field effects.

We uncover the intricate interplay between the angular momentum and many-body quantum correlations arising from strong interactions that govern both vortex formation and geometry - phenomena that extend beyond mean-field physics. The vortex geometry exhibits diverse configurations, including central, two, and three vortices and more complex structures like pentagons and diamonds. While the angular rotation generally enhances vortex production, interactions can facilitate or suppress this process. Notably, the corresponding mean-field predictions deviate significantly from the many-body results.

In the study of dynamics, we first initialize the system with a specific vortex structure. We suddenly reduce the interaction strength, revealing rich features in vortex post-quench dynamics. Each scenario of vortex dynamics involves two distinct timescales. The first timescale captures processes such as vortex breathing, distortion, revival, and pseudo-revival dynamics. The second timescale involves outward density modulation, which splits into several pieces equal to the initial number of vortices, independent of their arrangement. These fragments rotate in the opposite sense of the imprinting rotation. The vortex dynamics exhibit the same timescale as the dynamical fragmentation of the many-body wave function in single-particle orbitals for simple vortex structures, such as one or two vortices. In contrast, complex vortex structures display highly intricate dynamics, lacking a well-defined revival period.

The article is organized as follows. Section~\ref{sec:model_hamiltonian} presents the model Hamiltonian. In Section~\ref{sec:method}, we briefly present the methodology. Section~\ref{sec:initial_setup} addresses the initial configurations of vortices and fragmentation in the pre-quench state. Section~\ref{sec:dynamics} details the dynamics across several subsections, and Section~\ref{sec:conclusion} summarizes and concludes our findings.

\section{Model Hamiltonian}
\label{sec:model_hamiltonian}
We consider a 2D rotating BEC comprising $N$ atoms trapped in a disk potential with hard walls. The dynamics after an interaction quench is studied by solving the time-dependent Schr\"odinger equation,
\begin{equation}
\hat{H} \psi = i \hbar \frac{\partial \psi}{\partial t}.
\end{equation}

The many-body Hamiltonian in the rotating frame is given by

\begin{equation} 
\hat{H}(\mathbf{r}_1,\mathbf{r}_2, \dots \mathbf{r}_N)= \sum_{i=1}^{N} \hat{h}(\mathbf{r}_i) + \sum_{i<j}\hat{W}(|\mathbf{r}_i - \mathbf{r}_j|),
\label{propagation_eq}
\end{equation}
where $\hat{h}(\mathbf{r})$ is the one-body Hamiltonian,
\begin{eqnarray}
\hat{h}(\mathbf{r}) = &-&\frac{\hbar^2}{2m} \left( \frac{\partial^2}{ \partial x^2} + \frac{\partial^2}{\partial y^2} \right)-\frac{\hbar\Omega}{i} \left( x \frac{\partial}{\partial y} - y \frac{\partial} {\partial x} \right)\nonumber\\
&+& V_{pot} (\mathbf{r}).
\end{eqnarray}
Here, the first term is the kinetic energy, the second is related to the rotation, and the third is the trapping potential. The atoms are trapped by a disk potential in the $xy$ plane, 
\begin{eqnarray}
\label{eq:hardwall}
V_{pot}(r) = \begin{cases}
0, \quad & r \leqslant a, \\
V_0,  \quad & r > a, \\
\end{cases}
\end{eqnarray}
where $V_0>0$ is much larger than all other typical energy scales in the system. 

At ultracold temperatures, low-energy scattering provides an excellent description of the interparticle interactions. The specific shape of the two-body potential does not matter because the $s$-wave scattering length encapsulates all relevant information about the potential, allowing for a universal description of the scattering process, independent of its details~\cite{Bethe1949,Lima2023}. We assume a contact pairwise potential ${W}(|\mathbf{r}_i - \mathbf{r}_j|) = g \delta (|\mathbf{r}_i - \mathbf{r}_j|)$, where $g$ is the interaction strength. The delta function is implemented as a narrow and normalized Gaussian-shaped function,
\begin{equation}
    \label{eq:regulator}
    \delta_\Lambda(r) = \frac{ \Lambda^2\exp (-r^2\Lambda^2/2)} {2 \pi}.
\end{equation}
The function $\delta_\Lambda(r)$ is a regularization of the delta function, that is, smearing over distances $r<\Lambda^{-1}$, such that $\lim_{\Lambda\to\infty}\delta_\Lambda(r)=\delta(r)$~\cite{vanKolck2020}.

Hereafter, we report all results in dimensionless units. We scale distances in units of $L$, { $L$ being a typical length scale of the system}, and energy-related quantities in units of $\hbar^2/(m L^2)$. In the approach of Eq.~(\ref{eq:regulator}), $\Lambda$ should be large enough such that the results are independent of the regulator choice, which we have checked is true for our chosen value of $\Lambda=4$. Similarly, we have checked that Eq.~(\ref{eq:hardwall}) reproduces the expected hard wall potential results for $V_0=1000$.

\section{Method}
\label{sec:method}

We employ the multi-configurational time-dependent Hartree method for bosons~\cite{Streltsov:2006, Streltsov:2007, Alon:2007, Alon:2008} implemented in the MCTDH-X software package~\cite{Alon:2008,Lode:2016,Fasshauer:2016,Lin:2020,Lode:2020,MCTDHX}. As MCTDH-X solves the time-dependent many-body Schr\"odinger equation, it is the ideal method to probe quench dynamics of strongly interacting ultracold systems.

In this method, the many-body wave function $\Psi(t)$ is expanded as an adaptive superposition of all time-dependent permanents $\vert \mathbf{n};t\rangle$ as
\begin{equation}
\left| \Psi(t) \right>= \sum_{\mathbf{n}}^{} C_{\mathbf{n}}(t)\vert \mathbf{n};t\rangle.
\label{many_body_wf}
\end{equation}

The permanents are constructed by distributing $N$ bosons over $M$ self-consistent orbitals as
\begin{equation}
\vert \mathbf{n};t\rangle = \prod^M_{k=1}\left[ \frac{(\hat{b}_k^\dagger(t))^{n_k}}{\sqrt{n_k!}}\right] |0\rangle, 
\label{many_body_wf_2}
\end{equation}
where $\mathbf{n}=(n_1,n_2,...,n_M)$ represents the number of bosons in each orbital, $|0\rangle$ is the vacuum state, and $\hat{b}_k^\dagger(t)$ denotes the time-dependent operator that creates one boson in the $k$-th working orbital $\phi_k(\mathbf{r};t)$. The constraint $\sum_{k=1}^M n_k=N$ ensures the total number of bosons conservation. The distribution of $N$ bosons over $M$ orbitals, yields
\begin{equation}
\eta_p=\left(\begin{array}{c} N+M-1 \\ N \end{array}\right)
\end{equation}
number of permanents.

The expansion coefficients $C_{\mathbf{n}}(t)$ and the working orbitals $\phi_k (\mathbf{r}; t)$ are time-dependent and variationally optimized at every time step ~\cite{TDVM81}. This requires the stationarity of the action with respect to variations of the time-dependent coefficients and orbitals, resulting in a coupled set of equations of motion for these quantities, which are then solved simultaneously. Note that the one particle function { $\phi_k(\mathbf{r},t)$} and the coefficient $C_{\mathbf{n}}(t)$ are variationally optimal with respect to all parameters of the many-body Hamiltonian at any time~\cite{TDVM81,variational1,variational3,variational4}. Imaginary time propagation relaxes the system to its ground state, while real-time propagation provides the full dynamics of the many-body state.

The accuracy of the algorithm depends on the number of orbitals used. For $M=1$ (a single orbital), MCTDH-X reduces to the mean-field Gross-Pitaevskii approximation. The wave function becomes exact for $M \rightarrow \infty$ as the set $ \vert n_1,n_2, \dots ,n_M \rangle$ spans the complete $N$-particle Hilbert space. For practical calculations, we limit the number of orbitals to achieve convergence in relevant observables, which is confirmed when the occupation of the highest orbital becomes negligible.

We calculate several observables from the many-body state $|\psi(t) \rangle$ to extract relevant information. The degree of coherence is measured by the reduced one-body density matrix, defined as
\begin{eqnarray}
\rho^{(1)}(\mathbf{r},\mathbf{r}';t) &= \left< \Psi(t) \right| \hat{\Psi}^{\dagger}(\mathbf{r}) \hat{\Psi}(\mathbf{r}') \left| \Psi(t) \right> \nonumber \\
                    & = \sum_{k,q} \rho_{kq} \phi_k^{*}( \mathbf{r}',t )\phi_q(\mathbf{r},t),
\end{eqnarray}
where $\rho_{kq}= \langle \Psi | {\hat{b}}_k^{\dagger} {\hat{b}}_q|\Psi \rangle$, and $\hat{\Psi}(\mathbf{r}) = \sum_k  \phi_k ( \mathbf{r},t) \hat{b}_k(t)$, with $\hat{\Psi}^{\dagger}(\mathbf{r}) (\hat{\Psi}(\mathbf{r}))$  a field operator that creates (annihilates) a particle at position $\mathbf{r}$. The diagonal of $\rho^{(1)}(\mathbf{r},{\mathbf{r}}';t)$ corresponds to the one-body density.

The one-body reduced density matrix can be diagonalized to obtain its eigenfunctions, known as natural orbitals ${\phi}_{i}^{(NO)}(\mathbf{r},t)$, and their corresponding eigenvalues, which represent the orbital occupations $\rho_i$. The expansion of the one-body reduced density matrix in terms of the natural occupation and natural orbitals is given by
\begin{equation}
\rho^{(1)} (\mathbf{r}, \mathbf{r}';t) = \sum_i \rho_i \phi^{(\mathrm{NO}),*}_i(\mathbf{r}',t)\phi^{(\mathrm{NO})}_i(\mathbf{r},t).\label{eq:RDM1}
\end{equation}
During the time evolution of the system after the interaction quench, the orbital occupations provide the extent to which the Hilbert space is dynamically occupied.

Condensation and fragmentation are key many-body features that emerge from orbital occupation analysis. According to Penrose-Onsager criteria~\cite{Penrose}, the system of interacting bosons is considered condensed if a single natural orbital exhibits a macroscopic population, while it is fragmented when multiple orbitals have a macroscopic population~\cite{Muller}. We quantify the degree of fragmentation as
\begin{equation}
\label{eq:fragmentation}
F = 1 - n_1
\end{equation}
where $n_1$ is the largest natural occupation. It is well established that the degree of fragmentation depends on factors such as dimensionality, particle number, and interaction strength. However, it has been shown to universally decrease approximately as $1/\sqrt{N}$~\cite{Meso}.

\section{Initial setup: Vortex structure and fragmentation}
\label{sec:initial_setup}

\begin{figure}[!htb]
    \centering
    \includegraphics[width=0.50\textwidth]{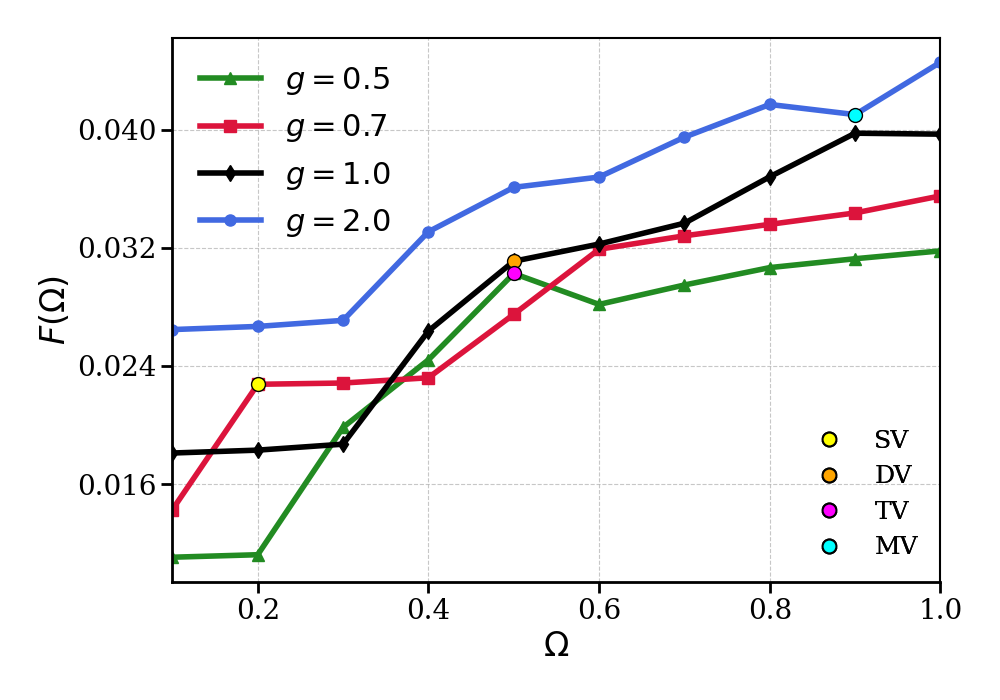}
    \caption{Degree of fragmentation $F$, Eq.~(\ref{eq:fragmentation}), as a function of the rotation frequency $\Omega$ and for interaction strengths $g$= $0.5, 0.7, 1.0, 2.0$. The highlighted points represent the different initial states indicated in Table \ref{tab:parameters}. }
    \label{fig1}
\end{figure}

\begin{figure}[!htb]
    \centering
    \includegraphics[width=\linewidth,angle=0]{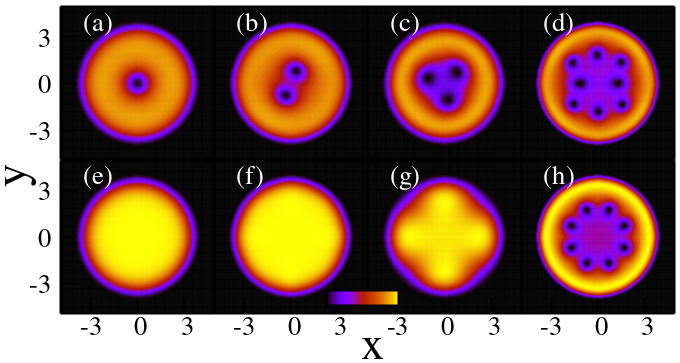}
    \caption{Initial one-body density distributions for four combinations of interaction strength $g$ and rotational frequency $\Omega$, summarized in Tab.~\ref{tab:parameters}, showing the generation of (a) a single vortex at the center, (b) a two-vortex structure, (c) a triangular vortex configuration, and (d) multiple vortices. The top panels (a)-(d) were obtained using the MCTDHB with $M=4$ self-consistent orbitals. The bottom panels (e)-(f) show corresponding mean-field results for the same parameters employing a single orbital.  }
    \label{fig2}
\end{figure}

The present work focused on understanding vortex stability and dynamics under an interaction quench of initial states. However, it is important to leverage the intricate interplay between angular frequency and interaction strength to obtain the desired vortex structure. First, we prepare the pre-quench state of $N=100$ bosons in the disk potential. All numerical simulations in this work employ 128 $\times$ 128 grid points, and the MCTDHB is used with $M=4$ self-consistent orbitals, compromising between increasing computational complexity (which scales
exponentially with the number of orbitals) and achieving orbital convergence.
We scan the rotation frequency $\Omega \in [0, 1]$ and interaction strength $g \in [0, 2]$. However, we consider four specific values of interaction strength $g= 0.5, 0.7, 1$, and $2$ covering the entire range of medium to strong interaction where the truly beyond mean-field effect is manifested. Figure~\ref{fig1} summarizes the fragmentation process. We present the degree of fragmentation $F$, Eq.~(\ref{eq:fragmentation}), across the entire range of rotational frequency for the previously mentioned choices of interaction strength. Figure~\ref{fig1} demonstrates that the rotating condensate remains fragmented across the entire range of parameters. However, unlike the nonrotating condensate,  we find the non-monotonic nature of fragmentation in the rotating condensate due to a strong interplay between interaction strength and rotational frequency. As expected, with increased interaction strength, fragmentation is stronger as more orbitals become populated, and $F$ increases with $g$. However, rotational frequency $\Omega$ plays an intricate role, and fragmentation passes through maxima and minima instead of increasing monotonically. To specifically configure the vortex structure, we scan it across the entire range of parameters as presented in Fig.~\ref{fig1}. However, for greater precision, we present specific cases in Table \ref{tab:parameters} that correspond to one, two, three, and multiple vortex structures as presented in Fig.~\ref{fig2}.

\setlength{\tabcolsep}{10pt}
\renewcommand{\arraystretch}{1.2}

\begin{table}[!htb]
\centering
\caption{Interaction strength $g$ and rotation frequency $\Omega$ used to generate the one, two, three, and multiple vortices configurations.}
\label{tab:parameters}
\begin{tabular}{|c|c|c|}
\hline
Configuration     & $g$ & $\Omega$ \\ \hline\hline
One vortex        & 0.7 & 0.2   \\ \hline
Two vortices      & 1.0 & 0.5   \\ \hline
Three vortices    & 0.5 & 0.5   \\ \hline
Multiple vortices & 2.0 & 0.9   \\ \hline
\end{tabular}
\end{table}

In Fig.~\ref{fig2}, we present the one-body density distributions for four different combinations of interaction strength and rotation frequency, as summarized in Table \ref{tab:parameters}, where the top panels correspond to many-body results, and the bottom panels are mean-field results. 
Figure~\ref{fig2}(a) corresponds to $g=0.7$ and $\Omega=0.2$, resulting in a single vortex at the center of the disk. Increasing both the interaction strength and rotation frequency ($g=1.0$ and $\Omega=0.5$) produces two vortices, as seen in Fig.~\ref{fig2}(b). The competition between the two parameters makes it so that if we keep the angular velocity at $\Omega=0.5$, then we have to reduce the interaction strength to $g=0.5$ to obtain a triangular vortex configuration, as depicted in Fig.\ref{fig2}(c). For a higher rotation frequency, $\Omega=0.7$, gradual tuning of the interaction strength reveals an evolution of vortex geometries. Starting from pentagonal structures, the vortex configuration transitions to a diamond shape, reverts to a triangular configuration, and finally reaches a two-vortex structure for stronger interactions (not shown here). Since we wanted to compare the few-vortex cases with a multiple-vortex configuration, we also chose to study relatively high rotation frequency and interaction strength, $\Omega=0.9$ and $g=2.0$, which produces a structure with eight vortices, as seen in Fig.~\ref{fig2}(d).

We observe vortex formation and geometry depend highly on angular frequency and interaction strength. While increasing angular velocity generally promotes the formation of vortices with diverse structures, the effect of interaction strength is more subtle. At moderate values, stronger interactions enhance vortex formation, whereas at relatively high interaction strengths, it suppresses and destroys the vortex structures (not shown here).

The bottom panels of Fig.~\ref{fig2} display the density profiles obtained from mean-field calculations for the same set of parameters used for the corresponding many-body calculations. They reveal that the mean-field approach fails to predict vortex formation in the first two cases, which correspond to one and two vortices according to the many-body simulations. Instead, the density profiles shown in Figs.~\ref{fig2}(e)-(f) exhibit a density profile resembling that of the non-rotating case. In Fig.~\ref{fig2}(g), it is possible to see that the mean-field simulation shows only a deformed cloud profile for the parameters corresponding to the three vortices case in the many-body calculation. In the multiple vortex case, Fig.~\ref{fig2}(h), the mean-field results predict eight vortices arranged in an octagonal configuration. In contrast, the many-body calculation yields the same number of vortices but distributed over a different configuration.

These discrepancies highlight a fundamental limitation of the mean-field theory, and the reason can be seen in Fig.~\ref{fig1}: the systems clearly manifest fragmentation, which plays a key role. Gradually increasing the $M$ values includes successive beyond mean-field effects, and fragmentation is related to orbital occupations. However, many-body orbitals are different from mean-field orbitals; density and correlations depend not only on occupations but also on many-body orbitals. Although the rotation frequency is small for the single vortex case ($g=0.7, \Omega=0.2$), the system exhibits strong fragmentation due to strong interaction and displays a local maximum in Fig.~\ref{fig1}. For the double and triple vortex cases, we keep the rotation frequency the same to some intermediate value $\Omega=0.5$, and fragmentation exhibits an intermediate peak in both cases, as shown in Fig.~\ref{fig1}. Many-body results significantly differ from mean-field results in these three cases, as shown in Fig.~\ref{fig2}, where fragmentation corresponds to a peak point. However, for the multiple vortex case ($g=2.0, \Omega=0.9$), fragmentation is a local minimum in Fig.~\ref{fig1}, resulting in a closer resemble of vortex structures obtained by mean-field and many-body formulations. It is to be noted that Fig.~\ref{fig1} demonstrates several other intermediate peaks and plateaus; the initial states can be configured in any of them. However, we configure the pre-quench states only for these four cases exhibiting different vortex geometries. The rest of the work is focused on the instability of vortex structures following an interaction quench.

\section{Dynamics following an interaction quench}
\label{sec:dynamics}

We explore the dynamical evolution of the vortices following a sudden interaction quench. The system is initially prepared in a particular vortex configuration as depicted in Fig.~\ref{fig2}. During the quenching process, the interaction strength is abruptly reduced to $g_i/100$, where $g_i$ is the initial interaction strength reported in Tab.~\ref{tab:parameters}. This induces vortex dynamics and subsequent instabilities that reveal unexplored many-body features in the post-quench states described in the following subsections. 

\begin{figure}[hbt]
    \centering
    \includegraphics[width=0.50\textwidth]{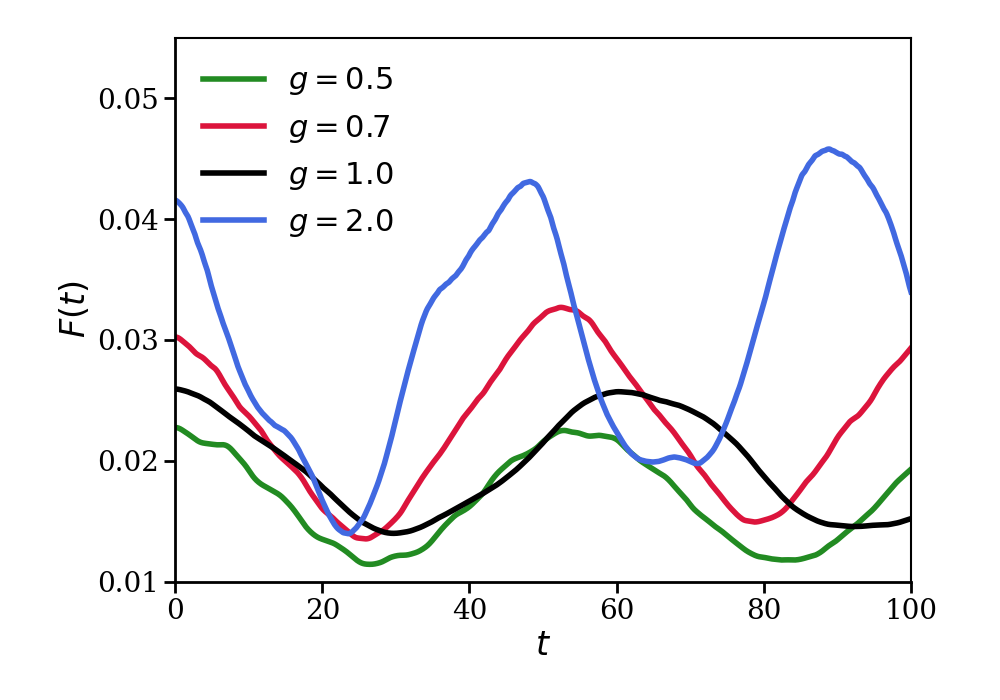}
    \caption{Degree of dynamical fragmentation in the sudden interaction quench dynamics for different initial states: one, two, three, and multiple vortex configurations. For each quench process, the interaction strength is suddenly reduced to one-hundredth of its initial value.}
    \label{fig3}
\end{figure}

Dynamical fragmentation plays a key role in determining the vortex dynamics. As described before, fragmentation is characterized by more than one significantly occupied orbital. The time evolution of natural occupation describes dynamical fragmentation. In the out-of-equilibrium dynamics of a non-rotating finite-sized trapped condensate, dynamical fragmentation strongly depends on interatomic correlations determined by the interaction strength, particle number, and elapsed time. There is a sharp contrast in the dynamic fragmentation between a quench consistent in a sudden increase or decrease of the interaction. In the first case, the fragmentation is enhanced as the less correlated state is quenched to a strongly correlated system. Initially, the lowest orbital is significantly occupied, and as time progresses, the population in it is reduced as the other orbitals become more populated. However, we should expect the opposite behavior in our quench corresponding to a sudden reduction of the interaction strength.

Figure~\ref{fig3} depicts the time evolution of the degree of fragmentation $F$, Eq.~(\ref{eq:fragmentation}), revealing the dynamical fragmentation during the post-quench dynamics. Initially, $F(t=0)$ starts from a maximum value, as the initial many-body system exists in a maximal fragmented state determined by the interplay of interaction strength and rotation frequency. In the sudden interaction quench, as the interaction is abruptly reduced, the system will lead to a less correlated and less fragmented state, indicating a gradual decrease in $F(t)$ with time. For all four configurations, we observe the same expected physics. It is also to note that $F(t=0)$ depicted in Fig.~\ref{fig3} ranges from $\sim$2\% to $\sim$4\%, indicating that the initial states are not very far from the non-fragmented condensate. However, following the discussion in Sec.~\ref{sec:initial_setup}, the dynamical evolution of many-body correlation will be strictly determined not only by time-dependent occupation in different orbitals but also by many-body orbitals, which are different from mean-field orbitals. These results point to the failure of mean-field theory to capture the many-body dynamics. Additionally, we observe some oscillatory patterns in $F(t)$. In the subsequent sections, we will discuss its relevance to the time scale of vortex dynamics.

The vortex dynamics comprises two different time scales; the first defines the dynamics of vortex structure, and the other determines the dynamics of the surrounding cloud density. The quench-induced time evolution reveals several intriguing phenomena, such as vortex expansion-contraction, distortion in vortex structure, combination, revival, and pseudo-revival. When vortices distort, irrespective of the initial vortex geometry, the outward cloud splits into the same number of fragments as the number of initial vortices. Conversely, the previously split clouds merge during the vortex revival or pseudo-revival dynamics. We divide our observations into four sections, each corresponding to a specific initial vortex geometry, as shown in Fig.~\ref{fig2}.

\subsection{Single vortex}

\begin{figure}[!htb]
    \centering
    \includegraphics[width=\linewidth,angle=0]{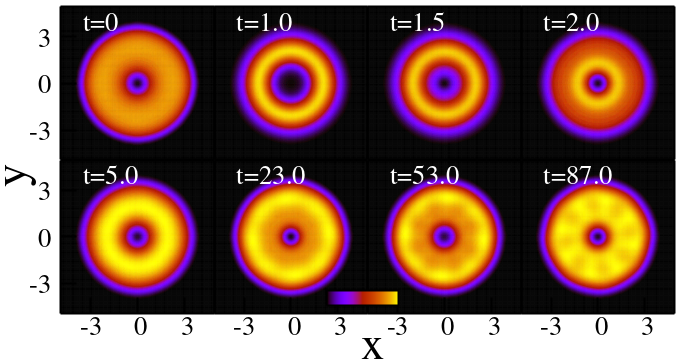}
    \caption{Snapshots of one-body density at selected times after an interaction quench for the single vortex case. The interaction strength is quenched to $g_f=0.007$ from $g_i=0.7$. The vortex dynamics primarily feature expansion-contraction cycles and revival phenomena.}
    \label{fig4}
\end{figure}

A single vortex is initialized at the center of the trap with an initial interaction strength $g_i=0.7$ and rotational frequency $\Omega=0.2$, as shown in Fig.~\ref{fig2}(a). The interaction strength is abruptly reduced to $g_f = 0.007$ during the quench. We have explored the dynamics for various values of $g_f$, and the key qualitative features remain robust. The evolution is studied up to $t=100$. Figure~\ref{fig4} shows snapshots of the post-quench density at selected times. The dominant behavior is a breathing mode, characterized by quasi-periodic expansion and contraction of the vortex core. Interestingly, the breathing dynamics are synchronized with the oscillations in dynamical fragmentation shown in Fig.~\ref{fig3}, with the vortex reaching its maximum and minimum size in phase with the fragmentation extrema.

\subsection{Two vortices}

\begin{figure}[!htb]
    \centering
    \includegraphics[width=\linewidth,angle=0]{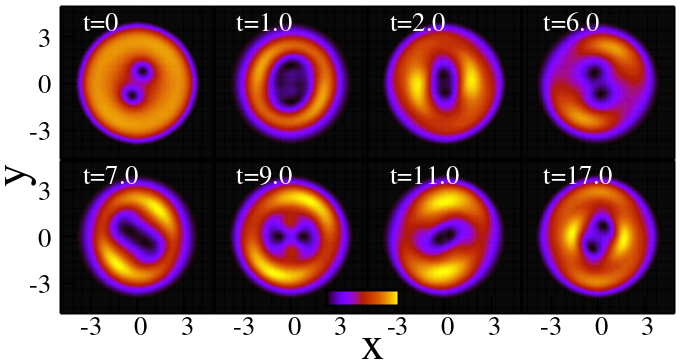}
    \caption{Snapshots of one-body density at selected times after an interaction quench for the double vortex case. The interaction strength is quenched to $g_f=0.01$ from $g_i=1.0$. The vortex dynamics exhibit distortion, merging, and pseudo-revival, with the outer cloud density splitting, rotating anticlockwise and eventually interacting at later times.}
    \label{fig5}
\end{figure}

As depicted in Fig.~\ref{fig2}(b), the initial double vortex structures correspond to interaction strength $g_i= 1.0$ and angular frequency $\Omega =0.5$. The interaction strength is suddenly reduced to $g_f=0.01$ during the quenching process. In Fig.~\ref{fig5}, we present snapshots of the one-body density for much shorter time intervals exhibiting intriguing features in the vortex dynamics. We observe two distinct types of dynamics; one is the distortion and evolution of the vortex structure itself, which we refer to as vortex dynamics, while the second concerns the rotational motion of the cloud surrounding the vortex.
 
As time progresses, we observe vortex recombination and the rotation of the density-split surrounded cloud in the opposite sense of the external rotational velocity. By $t=17.0$, two vortices appear, although not fully separated, and retain the same orientation as at $t=0.0$. However, the outward density cloud remains present, preventing a complete restoration of the original configuration, unlike the case of the single vortex. The corresponding dynamical fragmentation, shown in Fig~\ref{fig3}, exhibits oscillatory patterns similar to the single vortex case. With finer time resolution and long-time dynamics, we observe pseudo-revival phenomena coinciding with maxima and minima in the dynamical fragmentation. We conclude that the oscillatory nature of fragmentation is connected to the observed vortex dynamics; however, a strict one-to-one correspondence cannot be established.

\subsection{Three vortices}

\begin{figure}[!htb]
    \centering
    \includegraphics[width=\linewidth,angle=0]{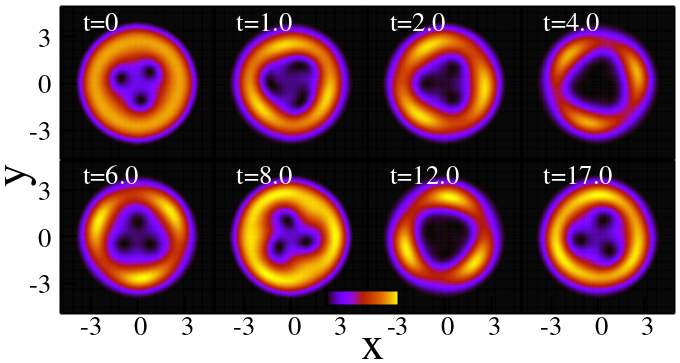}
    \caption{Snapshots of one-body density at selected times after an interaction quench for the triple vortex case. The interaction strength is quenched to $g_f=0.005$ from $g_i=0.5$. As the vortex structure merges and melts, the density cloud fragments into three distinct pieces, moving in an anticlockwise direction. These fragmented clouds interact and merge during the pseudo-revival of the vortex structure. The observed pseudo-revival dynamics align with the oscillatory behavior of dynamical fragmentation.}
    \label{fig6}
\end{figure}

The initial vortex state is prepared in a triangular geometry with $g_i= 0.5$ and $\Omega=0.5$ as depicted in Fig.~\ref{fig2} (c). In the post-quench dynamics, the interaction strength is suddenly reduced to $g_f=0.005$. The one-body densities at different times are depicted in Fig.~\ref{fig6}. We observe three distinct dynamical phenomena: deformation in the vortex core, the evolution of the outward density cloud, and the deformation of the rotating condensate. Between time $t=1.0$ to $t=3.0$, vortex distortion is observed, and the outward density cloud splits into three fragments that rotate in the opposite sense of the circulation of the vortices. At time $t=4.0$, the core structure is completely lost.  
 
At $t=8.0$, the vortex structure reappears with a different orientation, while the three split clouds merge into a single coherent ring structure. 

At $t$ =17.0, three vortices with the same orientation as the initial configuration revive, and the condensed outer cloud forms a ring surrounding the core of the three distinct vortices. The pseudo-revival phenomenon continues in the long-time dynamics that align with the oscillatory pattern observed in dynamical fragmentation, as shown in Fig.~\ref{fig3}. While we cannot establish exact correspondence between vortex structure collapse-revival and the extrema of dynamical fragmentation, the two phenomena are interrelated.

\subsection{Multiple vortices}

\begin{figure}[!htb]
    \centering
    \includegraphics[width=\linewidth,angle=0]{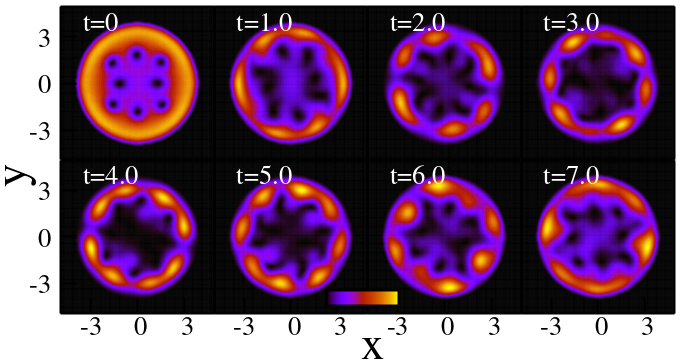}
    \caption{Snapshots of the one-body density evolution following an interaction quench in the multiple-vortex case. The interaction strength is quenched to $g_f=2.0$ from $g_i=0.02$. The time scale of the vortex dynamics is very fast compared to the previous cases. The dynamics show rapid deformation and reorientation of the vortex core, the fragmentation of the surrounding density cloud into eight segments, and a lack of vortex revival in the long-time dynamics.}
    \label{fig7}
\end{figure}

At larger angular velocity, the rapid rotation effectively destroys complex vortex configurations. With $\Omega=0.9$ and $g=2$, we create an initial vortex geometry of eight vortices arranged almost in a square. Intriguingly, instead of forming an octagonal arrangement, the vortices form a square geometry, suggesting that this square configuration with eight vortices minimizes the energy compared to an octagon. In the interaction quench process, we abruptly reduce the interaction strength to $g_f=0.02$. The corresponding density dynamics for consecutive time points are presented in Fig.~\ref{fig7}. 

At $t=1.0$, the initial square vortex structure begins to deform, and the outer density ring starts to split. By $t=5.0$, complete splitting occurs, revealing eight bright spots surrounding the central cloud.
As time progresses, the dynamics become increasingly chaotic. Unlike in previous cases, where pseudo-revival phenomena were observed, long-time evolution does not reveal any sign of revival of vortex structures.  
The dynamical fragmentation also exhibits an aperiodic oscillatory nature, as shown in Fig.~\ref{fig3}.

For rapid rotation, the initial vortex configurations calculated from many-body and mean-field methods show less substantial differences than in previously discussed cases. While both methods predict the same number of vortices, their arrangements differ [Fig.~\ref{fig2} (d) and (h)]. In contrast to the rectangular geometry observed in the many-body calculations, the mean-field theory predicts an octagonal arrangement for the eight vortices. This difference in vortex arrangement suggests that while the mean-field approach accurately captures the overall vortex count, it fails to fully account for the intricate energy minimization processes that govern vortex geometry in the many-body framework.

\begin{figure}[!htb]
    \centering
    \includegraphics[width=\linewidth,angle=0]{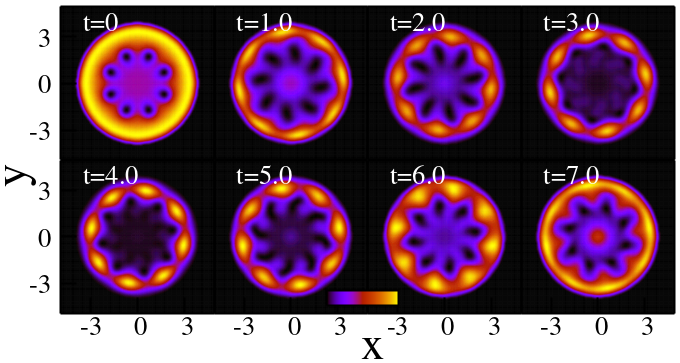}
    \caption{Snapshots of one-body density evolution following an interaction quench for the multiple-vortex case at the mean-field level, $M=1$. The interaction strength is quenched to $g_f=2.0$ from $g_i=0.02$. The vortex dynamics occur on a faster timescale compared to the previous cases. After the sudden quench, the inner core of the eight vortices adopts different orientations, with four vortices oriented upward and four downward. The surrounding ring cloud splits into eight distinct pieces. No vortex revival is observed in the long-time dynamics.}
    \label{fig8}
\end{figure}

After the interaction quench, the fundamental dynamical features are similar to those observed in the many-body dynamics. Figure~\ref{fig8} depicts the density dynamics at the same time points as the many-body results (Fig.~\ref{fig7}). At $t=1.0$, the eight vortices expand without merging, rotating counterclockwise, and the ring cloud splits into eight bright segments. However, unlike the many-body case, where the vortex structure becomes increasingly chaotic, the octagonal arrangement predicted by the mean-field theory remains intact. Throughout the evolution, the eight bright spots do not deform, in stark contrast to the many-body dynamics where vortices undergo significant rearrangement and eventual distortion. This difference highlights the limitations of the mean-field approximation in capturing the intricate many-body effects that govern vortex evolution under strong interactions. While the mean-field approach can qualitatively describe the overall expansion dynamics, it fails to replicate the rich, correlated behavior that emerges in the many-body simulations.

\section{Conclusion}
\label{sec:conclusion}

Rotating Bose-Einstein condensates confined in two-dimensional geometries have been extensively studied within ultracold atomic systems. While quantum vortices, their nucleation, and collapse are well understood within mean-field theory for weakly interacting systems, unprecedented experimental control now enables access to strongly interacting and ultrafast rotation regimes where the limitations of mean-field theory become apparent. Moving beyond mean-field approximations and exploring many-body effects is essential to understanding these systems fully.

In this work, we explored the vortex formation in a rotating Bose-Einstein condensate and their dynamics following interaction quench using {\it ab initio} numerical simulations of the many-body Schr\"odinger equation. Our many-body results extend the understanding of vortex dynamics in several key directions.

First, we demonstrate that a subtle interplay between angular velocity and interaction strength governs the formation and structure of vortices. Specifically, we focused on the strongly interacting regime where mean-field predictions deviate significantly from many-body results. The competition between interaction strength and rotation frequency creates single, double, triple, and multiple vortex structures. 

Second, we show that each vortex configuration undergoes a distinctive post-quench evolution, revealing rich many-body effects. The rotating condensate exhibits distinct yet synchronized, dynamical scales associated with the central vortices and the surrounding density cloud. The simplest case, single-vortex dynamics, primarily exhibits breathing motion with periodic expansion and contraction. The revival timescale of the vortex closely aligns with the oscillatory nature of dynamical orbital fragmentation. Double and triple vortex systems exhibit more complex behavior. As the vortex structures merge, the density cloud splits into two or three fragments, respectively. These split clouds undergo a rotating motion that reflects the circulation charge of all vortices. Pseudo-revivals are observed when the split clouds interact and merge, with the revival timescale closely related to the oscillations in the dynamical fragmentation. The dynamics become irregular in the case of multiple vortex structures, such as eight vortices arranged in a square geometry. The ring cloud splits into eight fragments, highlighting a universal feature of out-of-equilibrium vortex dynamics. Unlike simpler configurations, the eight-vortex system does not exhibit apparent revival behavior, and its long-time dynamics remain chaotic, as indicated by the aperiodic oscillations of the dynamical fragmentation.

Our study opens several potential avenues for future research. In our work, the circular symmetry of the system prevents any spatial density fragmentation. Future work could explore elongated traps where fragmentation due to rapid rotation induces density splitting, offering a new avenue to study the interplay between interaction, angular velocity, and fragmentation in asymmetric confinement. Furthermore, vortex dynamics in attractive Bose gases remain an underexplored domain, with existing studies largely restricted to weakly attractive condensates within the mean-field framework. Investigating strongly attractive systems could reveal novel phenomena and further challenge our understanding of vortex dynamics. In reference \cite{paolo25}, the authors investigate the importance of beyond-mean-field corrections in describing the long-range interactions in the vortex-rotating condensate. Investigating the impact of long-range interactions on the real-time dynamics of the vortices after the interaction quenches is another open avenue for research.

\section*{Acknowledgments}

This work was supported by the São Paulo Research Foundation (FAPESP) under the grants 2013/07276-1, 
2014/50857-8, 
2023/04451-9, 
2024/04637-8, 
2024/20641-5, 
and by the National Council for Scientific and Technological Development (CNPq) under grant 465360/2014-9. 
L.A.M acknowledges the support from Coordenação de Aperfeiçoamento de Pessoal de Nível Superior - Brasil (CAPES) - Finance Code 88887.999663/2024-00. 
Texas A\&M University is acknowledged.

\appendix

\section{Mean-field results for different system sizes}

\begin{figure}[!htb]
    \centering
    \includegraphics[width=\linewidth,angle=0]{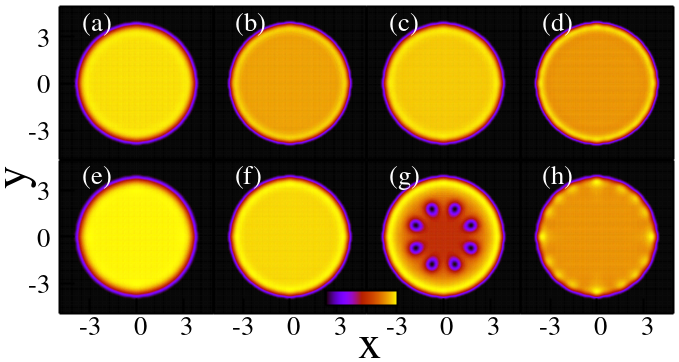}
    \caption{Initial one-body density distributions for four combinations of interaction strength $g$ and rotational frequency $\Omega$, summarized in Tab.~\ref{tab:parameters}, computed with $M=1$. (a) and (b) correspond to the parameters for a single vortex case with $N=500$ and $1000$, respectively. (c) and (d) correspond to the parameters for the double vortex case with $N=500$ and $1000$, respectively. (e) and (f) correspond to the parameters for the triple vortex case with $N=500$ and $1000$, respectively. (g) and (h) correspond to the parameters for the multiple vortex case with $N=200$ and $300$, respectively.}
    \label{fig9}
\end{figure}

In this Appendix, we generalize the bottom panel results of Fig.~\ref{fig2} from the main text for larger system sizes. Figures~\ref{fig9}(a) and (b) show the initial one-body density with $M=1$ from Fig.~\ref{fig2}(e) ($g=0.7, \Omega=0.2$) for $N=500$ and $1000$, respectively. No vortex is observed in the mean-field calculation, even with higher particle numbers. Figures~\ref{fig9}(c) and (d) show the case of Fig.~\ref{fig2}(f) with $N=500$ and $1000$ respectively, and demonstrate the same physics. Figures~\ref{fig9}(e) and (f), which correspond to Fig.~\ref{fig2}(g) with $N=500$ and $1000$, also fail to show any vortex structure, while the modulated central density is not visible with higher particle numbers. We can conclude that even in our simulation, finite-size effects do not dictate the main findings of this work. For multiple vortices, as shown in Fig.~\ref{fig2}(h), we repeat the calculation with several values of $N$ and plot two cases in Fig.~\ref{fig9}(g) ($N=200)$ and Fig.~\ref{fig9}(f) $(N=300)$. This shows that the vortex structure persists for $N=200$ but disappears for $N=300$. For simulations with higher particle numbers (not shown here), the vortex structure does not reappear. 

\section{Units and system parameters}

This section summarises the parameters used for numerical simulations in the main text. We have performed simulations with a fixed particle number $N=100$; the number of orbitals is $M=1$ for mean-field calculations, whereas, for many-body calculations, $M=4$ orbitals are used. 

The unit of length is $\bar{L}$. We perform the simulations in an interval $x, y \in [- 4 \bar{L}, 4 \bar{L}]$ with 128 grid points both in $x$ and $y$ directions. The unit of energy $\bar{E} \equiv \frac{\hbar^{2}}{m \bar{L}^{2}}$. The unit of time is defined in terms of the unit of length $\bar{t} \equiv \frac{m \bar{L}^{2}}{\hbar}$. Additionally, we work with natural units keeping $\hbar = m =1$.

\bibliography{ref}

\end{document}